# A generalization of Noether's theorem based on the virtual work functional

**D. H. Delphenich**[†]

In a series of previous articles by the author, it was shown that one could effectively give a variational formulation to non-conservative mechanical systems, as well as ones that subject to non-holonomic constraints by starting with the virtual work functional instead of an action functional. In this article, it is shown that when one starts with a virtual work functional that is not exact, so it does not admit an action functional, instead of conservation laws the extension of Noether's theorem gives balance principles for non-conserved currents. Examples of this generalized Noether identity are given in the mechanics of points, rigid bodies, and deformable bodies.

## 1  Introduction

Previously [**1, 2**], the author addressed the variational formulation of non-conservative systems, as well as systems with non-holonomic constraints, and showed that since the extremals of the action functional $S[x]$ are actually the zeroes of the first variation functional $\delta S[\delta x]$, it is more appropriate to regard that latter functional as the fundamental one. In fact, one can also interpret this functional as defining the total virtual work that is done by a virtual displacement.

The integrand of the virtual work functional is defined by a fundamental 1-form:

$$\phi = F_\mu \, dx^\mu + \Pi_\mu^a \, dx_a^\mu \tag{1.1}$$

on the manifold $J^1(\mathcal{O}; M)$ of 1-jets of local differentiable functions $x: \mathcal{O} \to M$; here, $\mathcal{O}$ is a $p$-dimensional parameter manifold with boundary and $M$ is an $m$-dimensional configuration manifold. The first term in $\phi$ represents the generalized forces that act on the system, while the second term represents the generalized momenta, which can also relate to infinitesimal stresses. One then regards the action functional as a special case scenario in which the fundamental 1-form $\phi$ is exact and representable as $d\mathcal{L}$ for some Lagrangian density function $\mathcal{L}$ on $J^1(\mathcal{O}; M)$.

The key to obtaining equations of motion from the virtual work functional is to recognize that it depends upon the vector field $\delta x$ on the image of $x$ by the intermediary of its 1-jet prolongation $\delta^1 x$; such a vector field on $J^1(\mathcal{O}; M)$ is then *integrable*. When one evaluates $\phi(\delta x)$ one then finds that precisely the same steps that one uses when $\phi = d\mathcal{L}$ lead to an integrand of the form:

$$D^*\phi(\delta x)\mathcal{V} = \left[\left(F_\mu - \frac{\partial \Pi_\mu^a}{\partial t^a}\right)\delta x^\mu\right]\mathcal{V} \tag{1.2}$$

that is integrated over $\mathcal{O}$ and another one that is integrated over $\partial\mathcal{O}$; With the usual boundary conditions on $\delta x$, this gives equations of motion in the form:

$$D^*\phi = \left(F_\mu - \frac{\partial \Pi_\mu^a}{\partial t^a}\right)dx^\mu = 0, \tag{1.3}$$

and when $\phi = d\mathcal{L}$, one finds that one has, indeed:

$$D^*\phi = \frac{\delta \mathcal{L}}{\delta x}, \tag{1.4}$$

since:

$$F_\mu = \frac{\partial \mathcal{L}}{\partial x^\mu}, \qquad \Pi_a^\mu = \frac{\partial \mathcal{L}}{\partial x_a^\mu} \tag{1.5}$$

when $\phi = d\mathcal{L}$.

---

[†] E-mail: david_delphenich@yahoo.com

One sees that, in effect, one is really using a generalization of d'Alembert's principle of virtual work (cf., e.g., Lanczos [**3**]) in place of Hamilton's principle of least action as the basis for the dynamical principle, since $D^*\phi(\delta x)$ takes the form of the virtual work that is done when the variation $\delta x$ is regarded as a virtual displacement of the object in $M$ that is described by the map $x$.

In this article, we first look at zeroes of the virtual work functional for more general variations of the kinematical state of the system that also include variations of the parameters in $\mathcal{O}$, as well. In order to be dealing with the usual machinery of Noether's theorem, one then considers the parameter space variations $\delta a$ to be infinitesimal generators $\mathfrak{a}$ of a Lie group action $G \times \mathcal{O} \to \mathcal{O}$. When one starts with an action functional, the Lie algebra homomorphism $\mathfrak{g} \to \mathfrak{X}(\mathcal{O})$, $\mathfrak{a} \mapsto J(\mathfrak{a})$ that Noether's theorem defines takes infinitesimal generators of the symmetries of the action functional to vector fields on $\mathcal{O}$ that have vanishing divergence; i.e., conserved currents. We shall find that when one starts with the virtual work functional instead of the action functional, this situation weakens only in that the resulting vector field $J(\mathfrak{a})$ does not generally have vanishing divergence. The resulting expression for its divergence then represents a balance principle, instead of a conservation law. For instance, in the case of point mechanics, it simply says that the current on $\mathcal{O} = [t_0, t_1]$ that is associated with time-translation invariance is kinetic energy, and its divergence – i.e., time derivative – equals the power delivered to or dissipated by the external forces that act on the moving mass.

In Section 2 of this article, we shall first summarize the basic results of the previous papers, as they relate to the problem at hand, and then present our generalization of Noether's theorem. In Section 3, we show how these general constructions specialize in the case where the virtual work functional is obtained from an action functional. In Section 4, we then define a level of specialization that is between the fully general virtual work functional and the exact ones that are defined by action functionals, namely, the case in which the momentum contribution to the fundamental 1-form $\phi$ is exact – viz., the differential of a kinetic energy function – even though the force contribution can still be inexact, as for non-conservative forces. In Section 5, we then discuss how the general results work in the context of the motion of point masses, rigid bodies, and deformable bodies. However, we shall only sketch the basic issues that are associated with deformable bodies, since that discussion would enlarge this study considerably. Finally, in the last section, we summarize the basic results of the paper.

## 2     The generalization of Noether's theorem

In order to properly explain the relationship between symmetry and conservation laws in the context of the variational calculus, one must start at a point that strictly precedes the starting point for the determination of extremals. This is because the class of variations that one deals with in obtaining the extremal equations does not generally include a contribution from the variation $\delta a^i$ of the independent parameters $a^i$, while the symmetries one considers in deriving Noether's theorem have infinitesimal generators that indeed represent variations of that nature.

### 2.1     Manifolds of 1-jets

More precisely, let $\mathcal{O}$ (for "object") be a compact, orientable, parameter manifold of dimension $p$ with a boundary $\partial \mathcal{O}$. Most commonly, when $\mathcal{O}$ is a static object, it will be a compact subset of $\mathbf{R}^p$ and its boundary will have one component, but when it is a dynamic object, $\mathcal{O}$ will take the form of $[t_0, t_1] \times \mathcal{O}_s$, where $\mathcal{O}_s$ is an open subset of $\mathbf{R}^{p-1}$ and the boundary $\partial \mathcal{O}$ consists of two components, namely, $\{t_0\} \times \mathcal{O}_s$ and $\{t_1\} \times \mathcal{O}_s$.

We shall refer to the coordinates of any local chart on $\mathcal{O}_s$ by the symbols $a_i$, $i = 1, \ldots, p$ and choose a volume element $\mathcal{V} \in \Lambda^p \mathcal{O}$ that has the local expression:

$$\mathcal{V} = da^1 \wedge \ldots \wedge da^p = \frac{1}{p!} \varepsilon_{i_1 \cdots i_p} da^{i_1} \wedge \cdots \wedge da^{i_p}. \tag{2.1}$$

Now let $M$ be an $m$-dimensional differentiable configuration manifold, which does not necessarily need to have a boundary or volume element of its own. We shall call any differentiable map $x: \mathcal{O} \to M$, $a \mapsto x(a)$ an *object* in $M$. Locally, when one chooses a local coordinate system $(U, a^i)$ on $\mathcal{O}$ and another one $(V, x^\mu)$ on $M$ the map $x$ can be expressed as a system of $m$ equations in $p$ independent variables:

$$x^\mu = x^\mu(a^i). \tag{2.2}$$

Actually, physical considerations suggest that we shall usually be thinking in terms of *embeddings*, since they represent the differentiable maps that preserve the topology of $\mathcal{O}$, if one realizes that the general differentiable map $x$ might very well map $\mathcal{O}$ to a lower-dimensional subset of $M$, such as a single point. Any embedding has the property that the differential map $dx|_a : T_a\mathcal{O} \to T_{x(a)}M$ to $x$ must have rank $p$ at every point $a \in \mathcal{O}$. More generally, if $x$ has this property one only calls it an *immersion*, the difference between immersions and embeddings generally taking the form of self-intersections. Since we will be dealing locally with 1-jets of differentiable maps, the only thing that will really matter is the condition on the rank of the differential map.

The *1-jet* $j_a^1 x$ of the object $x: \mathcal{O} \to M$ consists of three essential pieces of information: the point $a \in \mathcal{O}$, the corresponding point $x(a) \in M$, and the differential map $dx|_a: T_a\mathcal{O} \to T_{x(a)}M$, which can also be regarded as a tensor of mixed type in $T_a^*\mathcal{O} \otimes T_{x(a)}M$ that can be locally expressed as:

$$dx|_a = x_{,i}^\mu(a)\, da^i \otimes \partial_\mu. \tag{2.3}$$

One can then suitably generalize this notion, since when one is given only the pair of points $(a, x)$, without any specific functional relationship $x = x(a)$ in at least a neighborhood of each point, one cannot say whether a given element of $T_a^*\mathcal{O} \otimes T_x M$, which we express more generally by:

$$j_a^1 x = x_i^\mu\, da^i \otimes \partial_\mu, \tag{2.4}$$

represents the differential of some map $x: \mathcal{O} \to M$ at $a \in \mathcal{O}$. In general, it only represents an equivalence class of differentiable functions defined on various neighborhoods of $a$ that all take $a$ to the same point $x \in M$ and all have the same differential map at $a$.

The space of all these 1-jets of differentiable functions of $\mathcal{O}$ into $M$ is then a differentiable manifold $J^1(\mathcal{O}; M)$. The subset of all 1-jets of immersions is then a closed submanifold that represents a level hypersurface of the function on $J^1(\mathcal{O}; M)$ that takes each 1-jet to its rank, since all 1-jets of immersions will have rank $p$. A local coordinate system for $J^1(\mathcal{O}; M)$ will take the form $(a^i, x^\mu, x_i^\mu)$.

There are three canonical projections that one can consider for any $J^1(\mathcal{O}; M)$:

Source projection:    $\alpha: J^1(\mathcal{O}; M) \to \mathcal{O}$,    $j_a^1 x \mapsto a$,

Target projection: $\beta: J^1(\mathcal{O}; M) \to M$, $\quad j_a^1 x \mapsto x,$

Contact projection: $\pi_0^1: J^1(\mathcal{O}; M) \to \mathcal{O} \times M$, $\quad j_a^1 x \mapsto (a, x).$

Similarly, one can consider sections of each of these projections. We shall mostly be concerned with sections of the source projection, which then take the form of differentiable maps $s: \mathcal{O} \to J^1(\mathcal{O}; M)$, such that the projection of $s(a)$ is always $a$. They are represented in local coordinates in the form $(a^i, x^\mu(a), x_i^\mu(a))$. The reason for the choice of word "contact" in the third projection above is the fact that when $dx|_a$ has maximal rank, it takes the $p$-dimensional vector space $T_a\mathcal{O}$ to a $p$-dimensional vector subspace of the vector space $T_{x(a)}M$, which one refers to as a *contact element* at $x(a)$. A section of the contact projection is then referred to as a *field of contact elements* on the image of $\mathcal{O}$ in $M$; in the literature of geodesic fields [4], at least when $p = 1$, it takes the form of a "slope field."

A section $s$ of the source projection is called *integrable* iff it is the *1-jet prolongation* of a differentiable map $x: \mathcal{O} \to M$. Such a prolongation is defined by differentiation and takes the local form:

$$j^1 x(a) = (a^i, x^\mu(a), x_{,i}^\mu(a)), \tag{2.5}$$

in which the comma in the subscript refers to the partial derivative with respect to $a^i$.

Hence, one can say that $s$ is integrable iff:

$$s = j^1 x \tag{2.6}$$

for some $x: \mathcal{O} \to M$, which gives rise to the local condition:

$$x_i^\mu = x_{,i}^\mu, \tag{2.7}$$

which can also be regarded as a set of $pm$ first-order partial differential equations for the functions $x^\mu$.

### 2.2 Variations of kinematical states

We shall regard any section $s: \mathcal{O} \to J^1(\mathcal{O}; M)$, $a \mapsto s(a)$ as a *kinematical state* of the object $\beta \cdot s : \mathcal{O} \to M$. A *variation* of a kinematical state $s$ is vector field $\delta s(a)$ that is tangent to the image of $x$ in $J^1(\mathcal{O}; M)$. In terms of local coordinates it will take the form:

$$\delta s(a) = \delta a^i(a) \frac{\partial}{\partial a^i} + \delta x^\mu(a) \frac{\partial}{\partial x^\mu} + \delta x_i^\mu(a) \frac{\partial}{\partial x_i^\mu}. \tag{2.8}$$

A subset of these vector fields consists of 1-jet prolongations of variations of objects, which are, in turn, vector fields $\delta x(a)$ on $\mathcal{O} \times M$ that are tangent to the image of $x$. They then have the local form:

$$\delta x(a) = \delta a^i(a) \frac{\partial}{\partial a^i} + \delta x^\mu(a) \frac{\partial}{\partial x^\mu}. \tag{2.9}$$

The 1-jet prolongation $\delta^1 x(a)$ of a vector field $\delta x(a)$ on $\mathcal{O} \times M$ that is tangent to an object $x: \mathcal{O} \to M$ then has the local form:

$$\delta^1 x(a) = \delta a^i(a) \frac{\partial}{\partial a^i} + \delta x^\mu(a) \frac{\partial}{\partial x^\mu} + \frac{\partial(\delta x^\mu)}{\partial a^i}(a) \frac{\partial}{\partial x^\mu_i}, \tag{2.10}$$

A *dynamical state* for an object $\mathcal{O}$ is defined by a linear functional $\phi$ on the tangent space to $j_a^1 x \in J^1(\mathcal{O}; M)$; that is, it is a covector or 1-form. The dynamical state that is associated with the kinematical state $s(a)$ is then the pull-down $s^*\phi$ to a 1-form on $\mathcal{O}$. It basically says how the kinematical state $s$ will respond to a variation $\delta s$ by associating a differential increment of *virtual work* $\delta W = \phi(\delta s)$ with $\delta s$, which is regarded as a *virtual displacement*. We assume that $\phi$ is represented by a global 1-form $\phi$ on $J^1(\mathcal{O}; M)$ that has the local form:

$$\phi = F_\mu\, dx^\mu + \Pi^i_\mu dx^\mu_i, \tag{2.11}$$

whose first term represents the virtual work done by generalized forces and whose second term represents the virtual work that is done by inertial forces as a result of generalized momenta.

The virtual work functional for a given object $x$ is then a natural outgrowth of d'Alembert's principle, as it assigns each variation $\delta s$ with total virtual work that would be associated with it when one integrates the increments $\delta W[\delta s]$ over $\mathcal{O}$:

$$W|_s[\delta s] = \int_{\mathcal{O}} s^*\left[i_{\delta s}(\phi \wedge \mathcal{V})\right] = \int_{\mathcal{O}} \left[s^*(\phi(\delta s))\mathcal{V} - (s^*\phi) \wedge \#\delta a\right]. \tag{2.12}$$

In the integrand, we have introduced notations for the pull-back of functions and differential forms on $J^1(\mathcal{O}; M)$ by the section $s: \mathcal{O} \to J^1(\mathcal{O}; M)$ and the Poincaré dual $\#\delta a = i_{\delta a}\mathcal{V}$ of the vector field $\delta a$ on $\mathcal{O}$. Their local forms are:

$$s^*\phi = \left(F_\mu x^\mu_{,i} + \Pi^j_\mu \frac{\partial x^\mu_j}{\partial a^i}\right) da^i, \quad s^*(\phi(\delta s)) = \left(F_\mu \delta x^\mu + \Pi^j_\mu \delta x^\mu_j\right), \tag{2.13}$$

and:

$$\#\delta a = \frac{1}{p!}(\varepsilon_{i_1\cdots\hat{i}\cdots i_p}\,\delta a^i)\,da^{i_1}\wedge\cdots\widehat{da^i}\wedge\cdots\wedge da^{i_p}, \tag{2.14}$$

in which the functions $F_\mu$ and $\Pi^i_\mu$ depend on $a$ by way of $s(a) = (a, x^\mu(a), x^\mu_i(a))$.

When the virtual displacement $\delta s$ is integrable, so $\delta s = \delta^1 x$, one can do an integration by parts (i.e., the product rule for differentiation in $\delta s$), and the integrand in (2.12) can be expressed as:

$$\delta W(\delta^1 x) = D^*\phi(\delta x) + d[\Pi^i(\delta x)\#\partial_i] - (j^1 x^*\phi) \wedge \#\delta a, \tag{2.15}$$

in which we are defining the set of $p$ 1-forms on $M$:

$$\Pi^i = \Pi^i_\mu dx^\mu, \tag{2.16}$$

and in which we have introduced the notation:

$$D^*\phi = \left(F_\mu - \frac{\partial \Pi^i_\mu}{\partial a^i}\right)dx^\mu. \tag{2.17}$$

As we have seen in previous articles, and will discuss in the next section, the operator $D^*$ generalizes the variational derivative operator in the sense that the object $x$ is extremal iff:

$$D^*\phi(\delta x) = 0 \tag{2.18}$$

for all variations $\delta x$ of $x$ that either vanish on $\partial\mathcal{O}$ or satisfy the transversality condition that $\Pi^i_\mu \delta x^\mu$ vanishes on $\partial\mathcal{O}$.

The extremal equations then take the form of a generalization of Newton's equations:

$$F_\mu = \frac{\partial \Pi^i_\mu}{\partial a^i}. \tag{2.19}$$

The operator $D^*$ also represents a sort of "transpose" to the operator on "vertical" vector fields on $J^1(\mathcal{O}; M)$:

$$DX = (X^\mu_{,i} - X^\mu_i)\frac{\partial}{\partial x^\mu} \otimes da^i, \tag{2.20}$$

which vanishes iff $X$ is integrable.

Hence, we see that the equations of motion, which represent a sort of balance principle for the dynamical state $\phi$, are essentially dual to the corresponding condition $\delta s = \delta^1 x$, which expresses the integrability of the variation $\delta s$ on $J^1(\mathcal{O}; M)$.

The variation $\delta x$ of the object $x$ is a *symmetry* of the virtual work functional if:

$$\delta W\big|_{j^1 x}[\delta^1 x] = 0. \tag{2.21}$$

When $x$ is an extremal, these symmetries include the vertical variations $\delta x$ that are tangent to $M$, but in order to be dealing with the same class of variations that are usually employed in order to derive Noether's theorem, one must consider variations of the form that we are currently using, namely, ones that project under the source projection to non-zero vector fields $\delta a$ on $\mathcal{O}$.

If $\delta W\big|_{j^1 x}[\delta^1 x]$ vanishes for an extremal $x$ then, from (2.15), we have the identity:

$$d[\Pi^i(\delta x)\#\partial_i] = (j^1 x^* \phi) \wedge \#\delta a. \tag{2.22}$$

Since the $\Pi^i(\delta x)$ are all just smooth functions on $\mathcal{O}$, we can move them inside the # and then, using the definitions:

$$\text{div} = \#^{-1} \cdot d \cdot \#, \qquad \Pi(\delta x) = \Pi^i(\delta x)\#\partial_i, \tag{2.23}$$

we can re-express (2.22) in the form:

$$\# \, \text{div}[\Pi(\delta x)] = (j^1 s^* \phi) \wedge \#\delta a. \tag{2.24}$$

Hence, this relation between the divergence of the vector field $\Pi(\delta x)$ on $\mathcal{O}$ and the pull-down of $\phi$ by the prolongation of the section $s$ is our generalization of the usual Noether current identity, as we shall establish in the next sub-section.

In local form, we see from (2.24) that when $\delta x$ is a symmetry of $\delta S|_x$ for an extremal $x$, it must satisfy the partial differential equation:

$$\frac{\partial}{\partial a^i}(\Pi^i_\mu \delta x^\mu) = \left( F_\mu x^\mu_{,i} + \Pi^j_\mu \frac{\partial x^\mu_{,j}}{\partial a^i} \right) \delta a^i. \tag{2.25}$$

A further application of the product rule puts this into the form:

$$\frac{\partial J^i}{\partial a^i}(\delta x) = \Phi_i \, \delta a^i, \tag{2.26}$$

with:

$$J^i(\delta x) \equiv \Pi^i_\mu \delta x^\mu - (\Pi^j_\mu x^\mu_{,j})\delta a^i, \qquad \Phi_i \equiv x^\mu_{,j}\left( F_\mu \delta^j_i - \Pi^j_{\mu,i} \right), \tag{2.27}$$

as long as one restricts oneself to vector fields $\delta a$ on $\mathcal{O}$ with vanishing divergence.

We shall find it more physically useful to split the right-hand side of (2.25) so that:

$$J^i(\delta x) \equiv \Pi^i_\mu \delta x^\mu - \tfrac{1}{2}(\Pi^j_\mu x^\mu_{,j})\delta a^i, \qquad \Phi_i \equiv F_\mu x^\mu_{,j} + \frac{1}{2}\left(\Pi^j_\mu \frac{\partial x^\mu_{,j}}{\partial a^i} - \frac{\partial \Pi^j_\mu}{\partial a^i} x^\mu_{,j}\right). \quad (2.28)$$

Now we see that instead of defining a conserved current $J = J^i \partial_i$ we have defined a *non-conserved* one, except that we also have a balance principle for what happens to non-conserved part of the divergence.

At any rate, the formula for $J^i$ shows that it defines a linear map $J: \mathfrak{X}(\mathcal{O}\times M) \to \mathfrak{X}(\mathcal{O})$, $\delta x \mapsto J(\delta x)$, and if it is evaluated for $\delta s$ tangent to an extremal object $x$ then the vector field $J(\delta x)$ on $\mathcal{O}$ must satisfy the identity (2.26). Since the map $J$ takes one Lie algebra to another one, one naturally wishes to know whether it is also a Lie algebra homomorphism; that is, does one always have $[J(\delta x), J(\delta x')] = J[\delta x, \delta x']$? However, one rapidly discovers upon evaluating the two expressions in this equality that the components of the matrix of $J$ enter quadratically on the left-hand side, but linearly on the right-hand side.

Generally, one considers variations $\delta x$ whose components over $M$ consist of two terms:

$$\delta x^\mu = x^\mu_{,i} \delta a^i + \delta \overline{x}^\mu, \quad (2.29)$$

the first of which represents the push-forward of the vector field $\delta a$ on $\mathcal{O}$ by the map $x: \mathcal{O} \to M$, and the second of which represents the "substantial" (or essential) part of the variation.

When one inserts this into $J^i(\delta x)$, one finds that:

$$J^i(\delta x) = T^i_j \delta a^j + \Pi^i_\mu \delta \overline{x}^\mu, \quad (2.30)$$

in which we have introduced the generalization of the canonical stress-energy-momentum tensor:

$$T^i_j = \Pi^i_\mu x^\mu_{,j} - \tfrac{1}{2}(\Pi^k_\mu x^\mu_{,k})\delta^i_j. \quad (2.31)$$

This is where we can justify our previous choice of definitions for $J^i$ and $\Phi_i$, since if we take the trace of $T^i_j$ we get:

$$T = T^i_i = (1 - p/2)\Pi^i_\mu x^\mu_{,i}. \quad (2.32)$$

Had we chosen the definitions (2.27), this trace would always be zero, whereas we would prefer that it represent the generalized kinetic energy of the object, in some sense.

### 2.3 Generalized Noether theorem

One must note that so far it has not been necessary to specify that there was any sort of Lie group action on $\mathcal{O}$ or $M$. Hence, one can think of the classical Noether theorem as more of a corollary to a more general theorem that is expressed by equation (2.26).

In order set up the machinery for this, we need only suppose that a Lie group $G$ acts smoothly on $\mathcal{O} \times M$; that is, there is a smooth map $G \times (\mathcal{O} \times M) \to \mathcal{O} \times M$, $g(a, x) \mapsto (ga, gx)$, such $g'(g(a, x)) = (g'g)(a, x)$ and $e(a, x) = (a, x)$. This action differentiates at $e \times (a, x)$ to a linear map from $\mathfrak{g} \times T_{(a, x)}(\mathcal{O} \times M)$ to $T_{(a, x)}(\mathcal{O} \times M)$. For each $\mathfrak{a} \in \mathfrak{g}$, we shall call the tangent vector that $(\mathfrak{a}, 0)$ maps to under this map $\tilde{\mathfrak{a}}(a, x)$, which defines a vector field on $\mathcal{O} \times M$ that one calls the *fundamental vector field* associated with $\mathfrak{a} \in \mathfrak{g}$. One can also obtain it by exponentiating $\mathfrak{a}t$ to a one-parameter subgroup $\exp(\mathfrak{a}t)$ of $G$, letting this act on $(a, x)$, and then differentiating the resulting smooth curve in $\mathcal{O} \times M$ by $t$ at $t = 0$:

$$\tilde{\mathfrak{a}}(a, x) = \frac{d}{dt}\bigg|_{t=0} \exp(\mathfrak{a}t)(a, x) = \tilde{\mathfrak{a}}^i \frac{\partial}{\partial a^i} + \tilde{\mathfrak{a}}^\mu \frac{\partial}{\partial x^\mu}. \tag{2.33}$$

This means that to each $\mathfrak{a}$ in the Lie algebra $\mathfrak{g}$, there is a corresponding vector field $\tilde{\mathfrak{a}}$ in the Lie algebra $\mathfrak{X}(\mathcal{O} \times M)$. Furthermore, the map from $\mathfrak{g}$ to $\mathfrak{X}(\mathcal{O} \times M)$ that is thus defined is a Lie algebra homomorphism; i.e., $[\mathfrak{a}, \mathfrak{b}]$ goes to $[\tilde{\mathfrak{a}}, \tilde{\mathfrak{b}}]$. Under the subsequent linear map defined by $J: \mathfrak{X}(\mathcal{O} \times M) \to \mathfrak{X}(\mathcal{O})$, one then obtains a linear map from $\mathfrak{g}$ to $\mathfrak{X}(\mathcal{O})$. The generalization of Noether's theorem is then:

**Generalized Noether theorem**: *When the fundamental vector fields of a group $G$ that acts on $\mathcal{O} \times M$ are all symmetries of the first variation functional, every infinitesimal generator of $G$ – i.e., every element of $\mathfrak{g}$ – is associated with a vector field $J(\tilde{\mathfrak{a}})$ whose divergence satisfies* (2.26) *with $\delta a^i = \tilde{\mathfrak{a}}^i$.*

Since we can decompose the Lie algebra homomorphism $\mathfrak{g} \to \mathfrak{X}(\mathcal{O} \times M)$, $\mathfrak{a} \mapsto \tilde{\mathfrak{a}}$, into $\tilde{\mathfrak{a}} = \mathfrak{D}(\mathfrak{a}) + \overline{\mathfrak{D}}(\mathfrak{a})$ one part $\mathfrak{D}: \mathfrak{g} \to \mathfrak{X}(\mathcal{O})$, $\mathfrak{a} \mapsto \delta a(\mathfrak{a})$, and another part $\overline{\mathfrak{D}}: \mathfrak{g} \to \mathfrak{X}(M)$, $\mathfrak{a} \mapsto \delta \overline{x}(\mathfrak{a})$, we can set:

$$\delta \overline{x}^\mu(\mathfrak{a}) = \overline{\mathfrak{D}}^\mu_A \mathfrak{a}^A, \qquad \delta a^i(\mathfrak{a}) = \mathfrak{D}^i_A \mathfrak{a}^A, \tag{2.34}$$

which makes:

$$\Pi^i_\mu \delta \overline{x}^\mu = S^i_A \mathfrak{a}^A, \tag{2.35}$$

in which we have defined the *canonical spin tensor:*

$$S_A^i = \Pi_\mu^i \overline{\mathfrak{D}}_A^\mu. \tag{2.36}$$

When compared to the canonical stress-energy-momentum tensor, one sees that only the canonical spin tensor depends upon the nature of the action of *G* on *M*.

Hence, the Noether map from $\mathfrak{g}$ to $\mathfrak{X}(\mathcal{O})$ can be expressed by the matrix:

$$J_A^i = T_j^i \mathfrak{D}_A^j + S_A^i. \tag{2.37}$$

## 3      The case of exact virtual work functionals

Let us now return the more established methods of variational calculus, in which one starts with an action functional on objects *x*:

$$S[x] = \int_{\mathcal{O}} \mathcal{L}(j^1 x) \mathcal{V} \tag{3.1}$$

in which $\mathcal{L}: J^1(\mathcal{O}; M) \to \mathbf{R}$ is a differentiable function that one calls the *Lagrangian density* of the action functional.

One then finds that the process of establishing necessary and sufficient conditions for *x* to be an extremal first starts by deriving a first variation functional from $S[x]$ in which one basically arrives the association:

$$\phi = d\mathcal{L}; \tag{3.2}$$

i.e.:

$$F_\mu = \frac{\partial \mathcal{L}}{\partial x^\mu}, \qquad \Pi_\mu^i = \frac{\partial \mathcal{L}}{\partial x_i^\mu}. \tag{3.3}$$

That is, the first variation functional that one derives from an action functional takes the form of the virtual work functional.

One then finds that:

$$D^*\phi = \left( \frac{\partial \mathcal{L}}{\partial x^\mu} - \frac{\partial}{\partial a^i} \frac{\partial \mathcal{L}}{\partial x_i^\mu} \right) dx^\mu = \frac{\delta \mathcal{L}}{\delta x^\mu} dx^\mu, \tag{3.4}$$

and the extremal equations take the form of the Euler-Lagrange equations.

The issue of interest to us at the moment is how the exactness of $\phi$ affects the equations that we have derived for $J(\delta s)$. First, we go back to (2.12), and substitute from (3.3):

$$\delta W[\delta x] = j^1 x^* d\mathcal{L}(\delta^1 x) \mathcal{V} - d[d\mathcal{L}(j^1 x) \# \delta a]. \tag{3.5}$$

After integration by parts and assuming that $x$ is extremal, one is left with:

$$\delta W[\delta x] = d \,\#[\Pi(\delta x) - \mathcal{L}(j^1 x)\delta a]. \tag{3.6}$$

Hence, if $\delta x$ is a symmetry of $\delta S|_x$ for an extremal $x$, so the left-hand side of (3.6) vanishes, then one can define a conserved vector field:

$$J(\delta x) = \Pi(\delta x) - \mathcal{L}(j^1 x)\delta a = \left(\frac{\partial \mathcal{L}}{\partial x_i^\mu}\delta x^\mu - \mathcal{L}\delta a^i\right)\partial_i \tag{3.7}$$

that agrees with the usual Noether current.

When one decomposes $\delta x^\mu$ into its lifted and substantial pieces, the canonical stress-energy-momentum tensor takes the familiar form:

$$T_j^i = \frac{\partial \mathcal{L}}{\partial x_i^\mu} x_{,j}^\mu - \mathcal{L}\delta_j^i. \tag{3.8}$$

By contrast, the canonical spin tensor is not affected by the use of an exact fundamental 1-form $\phi$.

## 4 The case of exact kinetic work

A common situation in physical mechanics is when one might not have a conservative – i.e., exact – force term $F = F_\mu dx^\mu$ in $\phi$, but one does have an exact kinetic work term:

$$\Pi = \Pi_\mu^i dx_i^\mu = dT, \tag{4.1}$$

for some differentiable function $T = T(a^i, x^\mu, x_i^\mu)$ on $J^1(\mathcal{O}; M)$ that we regard as a generalized kinetic energy function for the system. One immediately sees that, as usual, such a function is not uniquely defined, although any two such functions will differ by a locally constant function (i.e., one that is constant on the connected components of $J^1(\mathcal{O}; M)$).

Since:

$$dT = \frac{\partial T}{\partial a^i}da^i + \frac{\partial T}{\partial x^\mu}dx^\mu + \frac{\partial T}{\partial x_i^\mu}dx_i^\mu, \tag{4.2}$$

an immediate consequence of the assumption that $\Pi$ has the form $dT$ is that:

$$\frac{\partial T}{\partial a^i} = 0, \qquad \frac{\partial T}{\partial x^\mu} = 0, \qquad \frac{\partial T}{\partial x_i^\mu} = \Pi_\mu^i. \tag{4.3}$$

Hence, $T$, as well as $\Pi^i_\mu$, can only be a function of $x^\mu_i$, so we also have:

$$\frac{\partial \Pi^l_\nu}{\partial a^i} = 0, \qquad \frac{\partial \Pi^l_\nu}{\partial x^\mu} = 0. \tag{4.4}$$

A necessary condition on $\Pi$ for it to be exact is that be closed:

$$0 = d\Pi = d\Pi^i_\mu \wedge dx^\mu_i = -\frac{1}{2}\left(\frac{\partial \Pi^i_\mu}{\partial x^\nu_j} - \frac{\partial \Pi^i_\nu}{\partial x^\nu_j}\right) dx^\mu_i \wedge dx^\nu_j; \tag{4.5}$$

of course, this is sufficient only if $J^1(\mathcal{O}; M)$ is simply connected, or rather, has vanishing de Rham cohomology in dimension one.

We define the fourth-rank tensor field on $J^1(\mathcal{O}; M)$ whose local components are:

$$\gamma^{ij}_{\mu\nu} = \frac{\partial \Pi^i_\mu}{\partial x^\nu_j}. \tag{4.6}$$

One sees that this tensor represents the constitutive law that associates generalized momenta with generalized velocities. The condition (4.5) can then be expressed as a symmetry property of this tensor:

$$\gamma^{ij}_{\mu\nu} = \gamma^{ji}_{\nu\mu}. \tag{4.7}$$

Hence, if one also assumes that the association of generalized velocities with generalized momenta is a linear isomorphism of the vector spaces of contact elements to $\mathcal{O} \times M$ with their dual spaces then we see that this symmetry property allows us to regard the tensor defined by the $\gamma^{ij}_{\mu\nu}$ as a scalar product on the spaces of generalized velocities by way of:

$$\gamma = \gamma^{ij}_{\mu\nu} dx^\mu_i dx^\nu_j. \tag{4.8}$$

### 4.1 Noether currents for exact kinetic work

Actually, from the standpoint of the virtual work functional the issue is whether the pull-down of $\Pi$ by $j^1 x: \mathcal{O} \to J^1(\mathcal{O}; M)$ is closed:

$$0 = d(j^1 x^* \Pi) = \left(\frac{d\Pi^k_\mu}{da^i} da^i\right) \wedge \left(\frac{dx^\mu_k}{da^j} da^j\right) = \tfrac{1}{2}[a^i, a^j]\, da^i \wedge da^j, \tag{4.9}$$

in which:

$$[a^i, a^j] = \frac{d\Pi_\mu^k}{da^i}\frac{dx_k^\mu}{da^j} - \frac{d\Pi_\mu^k}{da^j}\frac{dx_k^\mu}{da^i} \tag{4.10}$$

is the Lagrange bracket. Hence, the vanishing of this bracket is a necessary condition for the integrability of the kinetic work term.

If $\Pi = dT$ then we also have:

$$j^1x^*dT = d(j^1x^*T) = d(T(j^1x)), \tag{4.11}$$

and:

$$(j^1x^*\phi) \wedge \#\delta a = (j^1x^*F) \wedge \#\delta a + d(T(j^1x)) \wedge \#\delta a. \tag{4.12}$$

The second term becomes:

$$d(T(j^1x)) \wedge \#\delta a = d\#[T(j^1x)\delta a] = \# \text{div}[T(j^1x)\delta a] \tag{4.13}$$

which, upon substitution in (2.24) gives:

$$\#\text{div}[\Pi(\delta x) - T(j^1x)\delta a] = (j^1x^*F) \wedge \#\delta a. \tag{4.14}$$

Hence, the vector field $J$ on $\mathcal{O}$ that is associated with $\delta x$ is now:

$$J(\delta x) = \Pi(\delta x) - T(j^1x)\delta a = \left[\Pi_\mu^i \delta x_i^\mu - T(j^1x)\delta a^i\right]\frac{\partial}{\partial a^i}, \tag{4.15}$$

and although it is still not conserved, the right-hand side of the balance law:

$$\frac{dJ^i}{da^i}(\delta x) = F_\mu x_{,i}^\mu \delta a^i \tag{4.16}$$

now contains no contribution from $\Pi$ or its derivatives.

From a comparison of (4.15) with (2.28), we see that we can identify:

$$T(j^1x) = \tfrac{1}{2}\Pi_\mu^i x_i^\mu. \tag{4.17}$$

Similarly, a comparison of (4.16) with (2.28) shows that we must have:

$$\Pi_\mu^j \frac{dx_j^\mu}{da^i} = \frac{d\Pi_\mu^j}{da^i} x_j^\mu, \tag{4.18}$$

which can also be expressed as:

$$\Pi_\mu^j x_{j,i}^\mu = F_\mu x_j^\mu, \tag{4.19}$$

if one uses the equations of motion.

As for the canonical stress-energy-momentum tensor and the canonical spin tensor, they are now:

$$T^i_j = \frac{\partial T}{\partial x^\mu_i} x^\mu_{,j} - T\delta^i_j, \qquad S^i_A = \frac{\partial T}{\partial x^\mu_i} \mathfrak{D}^\mu_A. \tag{4.20}$$

The tensor $T^i_j$ differs from the Lagrangian expression by the absence of a contribution $+U\delta^i_j$ from a force potential $U(x)$.

### 4.2 Homogeneous kinetic energy

The most common form that kinetic energy takes, at least in point mechanics, is essentially:

$$T = \tfrac{1}{2} p(v) = \tfrac{1}{2} m \delta_{\mu\nu} v^\mu v^\nu, \qquad (p_\mu = m\delta_{\mu\nu} v^\nu). \tag{4.21}$$

Let us examine the consequences of assuming a generalization of this for our present $T$, namely:

$$T = \tfrac{1}{2} \Pi^k_\mu x^\mu_k. \tag{4.22}$$

Now, substitution of (4.4) in (4.22) makes:

$$T = \tfrac{1}{2} \frac{\partial \Pi^l_\nu}{\partial x^\mu_k} x^\mu_k x^\nu_l \equiv \tfrac{1}{2} \gamma^{kl}_{\mu\nu} x^\mu_k x^\nu_l. \tag{4.23}$$

In order for (4.23) to be satisfied, it is sufficient that $\Pi^k_\mu = \Pi^k_\mu(x^\nu_l)$ be homogeneous of degree one in $x^\mu_k$, which means that for any scalar $\lambda$, one has:

$$\Pi^k_\mu(\lambda x^\nu_l) = \lambda \Pi^k_\mu(x^\nu_l), \tag{4.24}$$

since Euler's theorem would then give:

$$\Pi^k_\mu = \frac{\partial \Pi^k_\mu}{\partial x^\nu_l} x^\nu_l. \tag{4.25}$$

As a consequence, the functions $\gamma^{ij}_{\mu\nu} = \gamma^{ij}_{\mu\nu}(x^\lambda_k)$ are then homogeneous of degree zero and the function $T = T(x^\mu_i)$ is homogeneous of degree two.

A function $T$ on $J^1(\mathcal{O}; M)$ that is homogeneous of degree one in the contact elements – i.e., the coordinates $x_i^\mu$ – can be the starting point for Finsler geometry [**5, 6**], at least if it is positive-definite. However, Finsler geometry is essentially a generalization of non-relativistic Lagrangian mechanics, which is not applicable at the moment as long as the force part of the fundamental 1-form $\phi$ is not assumed to also be exact.

## 5  Examples from physical mechanics

Since our discussion up to this point has been conspicuously lacking in physical examples, we shall rectify that oversight by showing how the general results specialize to the cases of the mechanics of point masses, rigid bodies, and deformable objects moving in space, although in the last case we shall only mention the basic issues.

### 5.1  Point mechanics

Let us first specialize our considerations to the most elementary case of point mechanics, in order to see what we are generalizing at that level.

The parameter manifold is one-dimensional ($p = 1$), namely, $\mathcal{O} = [t_0, t_1]$; it is then unnecessary to add the indices $i$, $j$, …, in the various expressions. The generalized velocity $x_i^\mu$ then takes the form of the usual velocity $v^\mu$, and the kinematical state space $J^1(\mathcal{O}; M)$ then becomes the manifold of all 1-jets of differentiable curve segments in $M$. Since a 1-jet of curves through a point is defined the same way as a tangent vector at that point, one finds that $J^1(\mathcal{O}; M) = [t_0, t_1] \times T(M)$. A section of the source projection $J^1(\mathcal{O}; M) \to [t_0, t_1]$ is then a differentiable curve in $T(M)$ – i.e., a vector field along a curve segment in $M$ – and a field of contact elements is a time-varying vector field on $M$:

$$\mathbf{v}(t, x) = v^\mu(t, x) \frac{\partial}{\partial x^\mu} ; \tag{5.1}$$

that is, the contact element is a tangent vector to $M$.

Similarly, a dynamical state in point mechanics is a 1-form $p$ on $[t_0, t_1] \times T(M)$ that annihilates vectors tangent to $[t_0, t_1]$, hence, it can be identified with a section of $[t_0, t_1] \times T^*M \to [t_0, t_1]$; i.e., a covector field along a curve in $M$:

$$p(t) = p_\mu(t)\, dx^\mu. \tag{5.2}$$

If we assume that $T(M)$ – and therefore $T^*M$ – has a metric $g$ defined on it, which may be either Riemannian or Lorentzian, then the usual way of relating the velocity of a curve in $M$ to its momentum is expressed locally by:

$$p_\mu = m g_{\mu\nu}\, v^\nu = (mv_0, mv_i), \tag{5.3}$$

in which $m > 0$ is a constant that can refer to either the mass in the non-relativistic (Riemannian) case or the rest mass in the relativistic (Lorentzian) one.

Of course, this relationship simplifies in an orthonormal local frame field to:

$$p_\mu = m\eta_{\mu\nu} v^\nu = (mv_0, -mv_i), \tag{5.4}$$

but in order for the orthonormal frame field to also be holonomic, like the natural frame fields that are defined by coordinate charts, the Levi-Civita connection that is defined by $g$ would have to be flat. Since the flat case still includes non-relativistic point mechanics, as well as special-relativistic point mechanics, we shall assume that case for the moment.

Note that $p_\mu$ is homogeneous of degree one in $v^\nu$. The kinetic energy that is then associated with $m$ and $g$ is then:

$$T(v) = \tfrac{1}{2} m\, g(v, v) = \tfrac{1}{2} mv^2, \tag{5.5}$$

at least, in the non-relativistic case.

In the relativistic case, the kinetic energy is not a Lorentzian frame-invariant function, but only one component of a Lorentzian frame covariant 1-form, namely, $p$. Indeed, the expression $mg(\mathbf{v}, \mathbf{v}) = p(\mathbf{v})$ becomes the rest energy $m_0 c^2$ for any physically admissible – i.e., proper-time parameterized – curve.

The dynamical part of the fundamental 1-form $\phi$ is the virtual work $F = F_\mu\, dx^\mu$ that is done by an (external) force whose components $F_\mu = F_\mu(t, x^\nu, v^\nu)$ then possibly vary with time, space, and velocity. One has:

$$\phi = F_\mu\, dx^\mu + dT. \tag{5.6}$$

As shown in [**1**], even when $F$ is not exact, as for a dissipative force, one can still obtain equations of motion from the vanishing of the first variation functional by way of:

$$0 = D^*\phi = \left(F_\mu - \frac{dp_\mu}{dt}\right) dx^\mu = \left(F_\mu - m\eta_{\mu\nu} \frac{dv^\nu}{dt}\right) dx^\mu, \tag{5.7}$$

namely, Newton's second law:

$$F_\mu = ma_\mu. \tag{5.8}$$

As for the balance law associated with time-translational symmetry, for which:

$$\delta a = \delta t, \qquad \delta x^\mu = v^\mu\, \delta t, \qquad \delta \bar{x} = 0, \tag{5.9}$$

we find that all that is left of the stress-energy-momentum $T^i_j$ tensor is:

$$T = p_\mu v^\mu - \tfrac{1}{2} p_\mu v^\mu = \tfrac{1}{2} p_\mu v^\mu, \tag{5.10}$$

so the generalized Noether current is:

$$J = T\, \delta t, \tag{5.11}$$

and the generalized Noether law is:

$$\frac{dT}{dt} = F_\mu v^\mu, \tag{5.12}$$

which expresses the fact that the time rate of change of kinetic energy along the curve of motion followed by the point mass $m$ equals the power that is delivered to (resp., dissipated from) $m$ due to the presence of the force $F$.

The requirement that one consider only variations $\delta t$ with vanishing divergence now takes the form of requiring that:

$$\frac{d(\delta t)}{dt} = 0, \tag{5.13}$$

which means that $\delta t$ can only represent constant time translations.

### 5.2    Rigid body

The next step in generality beyond the motion of a point mass, which is described by a differentiable curve $x(t)$ in a configuration manifold $M$, is the motion of a rigid body, which is described by an oriented orthonormal frame field along a differentiable curve in $M$. Although if all one wishes to do is discuss the rigid body then it is generally simpler to think of that curve as living in the group $ISO(3) = SO(3) \times_s \mathbf{R}^3$ of rigid motions of Euclidian $\mathbf{R}^3$, where the choice of the symbol $ISO(3)$ is based in the fact that the group is sometimes referred to as the inhomogeneous special orthogonal group, and the symbol $\times_s$ refers to the semi-direct product of the two groups.

#### 5.2.1    The Lie group $ISO(3)$ of rigid motions in space

One can define the semi-direct product directly by giving the Cartesian product of the sets the multiplication rule:

$$(R_1, a_1)(R_2, a_2) = (R_1 R_2, a_1 + R_1 a_2). \tag{5.14}$$

However, if one prefers to think of the action of linear groups on vector spaces, one can treat the coordinates $x^i$, $i = 1, 2, 3$ of $\mathbf{R}^3$ as really being the inhomogeneous coordinates of $\mathbf{RP}^3$ for some Plücker coordinate system and then embed $\mathbf{R}^3$ in $\mathbf{R}^4$, which represents the homogeneous coordinates of $\mathbf{RP}^3$, by way of the points of the form $(1, x^i)$. The group $ISO(3)$ can then be represented in $SL(4)$ by invertible 4×4 real matrices of the form:

$$g^\mu_\nu = (R, a) = \begin{bmatrix} 1 & | & 0 \\ \hline a^i & | & R^i_j \end{bmatrix} \qquad (5.15)$$

whose determinant is unity. One immediately verifies that matrix multiplication gives the same group multiplication rule as (5.14).

One the finds that by differentiating a curve $g(t) \in ISO(3)$ that the velocity vectors that are tangent to that curve consist of matrices of the form:

$$\dot{g}(t) = \begin{bmatrix} 0 & | & 0 \\ \hline \dot{a}^i(t) & | & \dot{R}^i_j(t) \end{bmatrix}. \qquad (5.16)$$

If one left-translates the points of the curve $g(t)$ back to the identity $e \in ISO(3)$ by way of $g^{-1}(t)$ and its tangent vectors $\dot{g}(t)$ by the differential of that left-translation then one obtains a curve in the Lie algebra $\mathfrak{gal}(3)$ of $ISO(3)$ of the form:

$$\omega^\mu_\nu(t) = g^{-1}(t)\dot{g}(t) = \begin{bmatrix} 0 & | & 0 \\ \hline \tilde{R}^i_j \dot{a}^j(t) & | & \omega^i_j(t) \end{bmatrix}, \qquad (5.17)$$

in which the tilde on $R$ signifies the inverse of that rotation matrix and:

$$\omega^i_j(t) = \tilde{R}^i_k(t) \dot{R}^k_j(t) = \begin{bmatrix} 0 & -\omega_z(t) & \omega_y(t) \\ \omega_z(t) & 0 & -\omega_x(t) \\ -\omega_y(t) & \omega_x(t) & 0 \end{bmatrix} \qquad (5.18)$$

is a curve in the Lie algebra $\mathfrak{so}(3)$ of infinitesimal generators of one-parameter subgroups of three-dimensional rotations. One can think of the matrix $\omega$ as the matrix ad($\omega$) of the adjoint map associated with the 3-vector $\boldsymbol{\omega} = (\omega_x, \omega_y, \omega_z)$ in $\mathbf{R}^3$, when it is given the Lie algebra structure of the vector cross product, which is the isomorphic to $\mathfrak{so}(3)$; i.e.:

$$\text{ad}(\boldsymbol{\omega})\mathbf{x} = \boldsymbol{\omega} \times \mathbf{x}. \qquad (5.19)$$

One can clearly think of the matrix $\omega^\mu_\nu(t)$ as consisting of the semi-direct product of the linear velocity $\dot{a}^i(t)$ of motion and the angular velocity $\omega^i_j(t)$.

The difference between dealing with curves $g(t)$ in $ISO(3)$ with tangent vectors of the elementary form $\dot{g}(t)$ and constant curves through its identity with tangent vectors of the form $\omega(t)$ amounts to the difference between describing the motion of a rigid body in an "inertial" frame and describing it in a "non-inertial" one that moves with the body itself.

### 5.2.2 The action of *ISO*(3) on *M* and kinematical states

Although it is entirely possible, and sometimes more desirable, to deal with *ISO*(3) as the configuration manifold of the rigid body (see, for instance, Arnol'd [7]), nevertheless, since we shall be taking the more physical viewpoint of regarding a rigid body as an approximation to a deformable extended body, it will be more convenient for us to anticipate the generalization from rigid to non-rigid extended matter by using the same formalism for both, namely, the geometry of jets.

First, we assume that the group *ISO*(3) of rigid motions acts smoothly on the three-dimensional manifold *M*, if only locally. That is, every point $x \in M$ has an open neighborhood *U* such that there is a smooth map $ISO(3) \times U \to M$, $(g, x) \mapsto gx$ such that:

$$g_2(g_1 x) = (g_2 g_1) x \text{ and } ex = x. \tag{5.20}$$

With no loss of generality, we can assume that *U* is homeomorphic to $\mathbf{R}^3$, so the action of *ISO*(3) on *U* is equivalent to an action on $\mathbf{R}^3$, which we assume takes the explicit form of:

$$\bar{x} = gx = (R, u)x = Rx + u. \tag{5.21}$$

If we pass from inhomogeneous coordinates to homogeneous ones then this can be represented by a linear action of *ISO*(3) on $\mathbf{R}^4$:

$$\bar{x}^\mu = g_\nu^\mu x^\nu = \begin{bmatrix} 1 & 0 \\ u^i & R^i_j \end{bmatrix} \begin{bmatrix} 1 \\ x^j \end{bmatrix} = \begin{bmatrix} 1 \\ u^i + R^i_j x^j \end{bmatrix}; \tag{5.22}$$

the reason that we are now representing the displacement vector by $u^i$ instead of $a^i$ is two-fold: We do not wish to confuse it with linear acceleration and we shall try to be more consistent with the notation of continuum mechanics for motions that include deformations.

Our way of representing the motion of a point in *M* will be to start with some initial point $x_0 \in M$ at $t = t_0$ and let a differentiable curve $g(t)$ in *ISO*(3) act on it:

$$x(t) = g(t) x_0. \tag{5.23}$$

Note the subtlety associated with the fact that we have replaced the six-dimensional group *ISO*(3) with the infinite-dimensional group $C^1([t_0, t_1], ISO(3))$ of differentiable curve segments in it.

By differentiation, we obtain the velocity vector field $v(t)$ to the curve $x(t)$ in the form:

$$v(t) = \dot{g}(t) x_0 + g(t) v_0, \tag{5.24}$$

in which $v_0$ is the initial velocity vector at $t_0$.

We can now introduce two types of kinematical states into this picture: 1-jets of differentiable curves in *M*, such as $(t, x, v)$, and 1-jets of differentiable curves in *ISO*(3), such as $(t, g, \dot{g})$. Thus, we need to consider sections of the source projections of both $J^1([t_0, t_1]; M)$ and $J^1([t_0, t_1]; ISO(3))$, which will then take the forms $(t, x(t), v(t))$ and $(t, g(t), \dot{g}(t))$, respectively.

We then see that the action $ISO(3) \times U \to M$ prolongs to an action $J^1([t_0, t_1]; ISO(3)) \times J^1([t_0, t_1]; U) \to J^1([t_0, t_1]; M)$, which takes the pair $(t, g, \dot{g}) \times (t_0, x_0, v_0)$ to the jet $(t, x, v)$, with:

$$x = gx_0, \qquad v = \dot{g}x_0 + gv_0. \tag{5.25}$$

Note that we are not dealing with sections now, but only points in the two jet manifolds.

So far, we have been discussing motion with respect to an inertial frame. In order to discuss motion with respect to non-inertial frames, we then have to change our definition of a kinematical state in $J^1([t_0, t_1]; ISO(3))$ and its action on the kinematical states in $J^1([t_0, t_1]; U)$. When everything in $ISO(3)$ gets left-translated to the identity, a kinematical state in $J^1([t_0, t_1]; ISO(3))$ will take the form $(t, e, \omega)$. As a consequence, the integrable section $j^1 g(t) = (t, g(t), \dot{g}(t))$ of the source projection of $J^1([t_0, t_1]; ISO(3))$ will be associated with a non-integrable one $s_g(t) = (t, e, \omega(t))$. It is precisely this non-integrablility that defines the soul of rotational mechanics.

The action of $(t, e, \omega)$ on $(t_0, x_0, v_0)$ is then defined by factoring $g$ out of (5.25) and setting $\omega = g^{-1}\dot{g}$:

$$\bar{x}_0 \equiv g^{-1}x = x_0, \qquad \bar{v}_0 \equiv g^{-1}v = \omega x_0 + v_0. \tag{5.26}$$

### 5.2.3 Moving frames as jets

Actually, the picture that we have defined so far for the motion of a rigid is incomplete, since we are approximating the extended body $B \subset \mathbf{R}^3$ by a single point, such as the origin in the parameter space $\mathbf{R}^3$, which we will assume coincides with the center of mass of the mass distribution in $B$. However, in order to physically account for the rotation of $B$ in time, one also needs to associate the rigid body with an oriented, orthonormal frame, such as $\mathbf{e}_i(t) = g_i^j(t)\partial_j$, $i = 1, 2, 3$ with $g_i^j(t) \in SO(3)$, at the center-of-mass. In order to obtain an oriented, orthonormal frame field along a differentiable curve in $M$ it would help to have a linear isomorphism $\mathbf{e}: \mathbf{R}^3 \to T_{x(t)}M$, $\partial_i \mapsto \mathbf{e}(t)$ for each $t \in [t_0, t_1]$; in fact, this is one way of defining a linear frame in $T_{x(t)}M$. Now, if we were looking at differentiable maps $x: [t_0, t_1] \times B \to M$ that were diffeomorphisms of $B$ with its image in $M$ for each $t$ then this linear isomorphism could be obtained from $dx|_{(t, 0)}$. Hence, a linear frame $\{\mathbf{e}_i, i = 1, 2, 3\}$ in a tangent space $T_xM$ can be associated with the 1-jet $j^1 x = (t, a^j, x^i, v^i, x_j^i)$ of a differentiable map $x: [t_0, t_1] \times B \to M$, $(t, a) \mapsto x(t, a)$ that is a diffeomorphism of $B$ onto its image in $M$ for each $t$ when one restricts the domain of $x$ to $[t_0, t_1] \times \{0\}$. Therefore, the approximation of an extended body by a point is not a topological *contraction* that reduces the dimension, but simply the *restriction* of the map that defines the motion of the extended object $B$ to one that follows the motion of one of its specified points, such as the center of mass. However, the association of a moving frame along the curve $x(t) = x(t, 0)$ still depends upon the assumption that $B$ has finite spatial extent.

A section of $J^1([t_0, t_1] \times B; M) \to [t_0, t_1] \times B$ will take the form:

$$s(t, a) = (t, a, x^i(t, a), v^i(t, a), x^i_j(t, a)). \tag{5.27}$$

When restricted to $a = 0$ this can be abbreviated to:

$$s(t) = (t, x^i(t), v^i(t), x^i_j(t)), \tag{5.28}$$

which differs from a section of $J^1([t_0, t_1]; M) \to [t_0, t_1]$ only by the addition of the final coordinates $x^i_j(t)$, which represent the linear frame in $T_{x(t)}M$ by way of:

$$\mathbf{e}_j(t) = x^i_j(t) \frac{\partial}{\partial x^i}. \tag{5.29}$$

In order to be speaking of oriented, orthonormal frames instead of more general linear ones, one need only restrict oneself further to matrices $x^i_j$ that belong to the special orthogonal group $SO(3)$. Thus, we are really concerned with only a submanifold of $J^1([t_0, t_1] \times B; M)$, not the entire manifold.

One sees that we are still kinematically incomplete in the eyes of rotational mechanics, since we need to account for the time rate of change of $\mathbf{e}_j(t)$, as well as its angular orientation. When one adds the components $v^i_j$ of the time derivative of $x^i_j$ to the kinematical state it takes the form $(t, x^i, v^i, x^i_j, v^i_j)$ and rearranges this to $(t, x^i, x^i_j, v^i, v^i_j)$, one sees that it is more straightforward to regard a kinematical state of a rigid body in $M$ as a 1-jet of a differentiable curve in the bundle $SO(M)$ or oriented, orthonormal 3-frames in it, since locally $SO(M)$ looks like $(x^i, x^i_j)$, $x^i_j \in SO(3)$.

Therefore, we now need to extend the action of a 1-jet of the form $(t, g, \dot{g})$ or $(t, e, \omega)$ to the 1-jets of the form $(t, x^i, x^i_j, v^i, v^i_j)$. Since we already have the action defined on $(t, x^i, v^i)$ in either case, we need only add the action on $x^i_j$ and $v^i_j$. This is obtained by differentiating $x(t, a) = g(t)x_0(a)$ with respect to the spatial variables $a^i$ and generalizing the resulting expression:

$$x^i_j = g[x^i_j]_0, \tag{5.30}$$

which also makes:

$$v^i_j = \dot{g}[x^i_j]_0 + g[v^i_j]_0. \tag{5.31}$$

Thus, we now have the action of the 1-jets in $J^1([t_0, t_1]; ISO(3))$ on the 1-jets in $J^1([t_0, t_1]; SO(U))$.

We need to point out that we still need to extend the action of *ISO*(3) on the frames of *SO*(*M*), as well as on the points of *M* in order to fully extend our previous formalism from the manifold *M* to the manifold *SO*(*M*). However, this is straightforward if one agrees to project *ISO*(3) onto *SO*(3) by taking each $g_\nu^\mu \in$ *ISO*(3) to its rotational submatrix $R_j^i \in$ *SO*(3), which then acts naturally on oriented, orthonormal 3-frames by matrix multiplication. From (5.31), we see that this has the effect of introducing two types of angular velocity: one of them, $\dot{g}[x_j^i]_0$, is due to the "orbital" rotation of points in *M*, while the other, $g[v_j^i]_0$, is due to the "intrinsic" rotation of the frame about its origin.

In order to describe the action of ($t$, $e$, $\omega$) on ($t$, $x^i$, $x_j^i$, $v^i$, $v_j^i$), we need only to transform (5.30) and (5.31) back to the initial frame:

$$[\bar{x}_j^i]_0 \equiv g^{-1} x_j^i = [x_j^i]_0, \qquad [\bar{v}_j^i]_0 = g^{-1} v_j^i = \omega[x_j^i]_0 + [v_j^i]_0. \tag{5.32}$$

We see that this is indeed the frame in which the moving frame itself appears to stand still, but an initial intrinsic angular velocity $[v_j^i]_0$ for some rotational motion that is being viewed in that moving frame has acquired an additional orbital angular velocity $\omega[x_j^i]_0$ due to the fact the frame of reference is itself rotating with respect to an inertial frame. For instance, if one is observing the second hand on a horizontal watch while rotating on a merry-go-round then this expression would be applicable.

### 5.2.4 Dynamical states of the moving rigid body

If our kinematical states are elements of $J^1([t_0, t_1]; SO(M))$ then our dynamical states should be 1-forms $\phi$ on $J^1([t_0, t_1]; SO(M))$ that have the local form:

$$\phi = F_i \, dx^i + \tau_i^j dx_j^i + p_i \, dv^i + S_i^j dv_j^i. \tag{5.33}$$

In order to convert this into the form $\phi_g + \phi_0$, where $\phi_g$ is a 1-form on $J^1([t_0, t_1]; ISO(3))$ that looks like:

$$\phi_g = M_i^j dg_j^i + L_i^j d\dot{g}_j^i \tag{5.34}$$

in an inertial frame, and $\phi_0$ is a 1-form on $J^1([t_0, t_1]; SO(U))$ that looks like:

$$\phi_0 = F_{0i} dx_0^i + p_{0i} dv_0^i + [\tau_i^j]_0 d[x_j^i]_0 + [S_i^j]_0 d[v_j^i]_0, \tag{5.35}$$

we need to differentiate the group action (4.22), (4.27), (4.28):

$$x = gx_0, \qquad v = \dot{g}x_0 + gv_0, \qquad e = ge_0, \qquad \dot{e} = \dot{g}e_0 + g\dot{e}_0. \tag{5.36}$$

in which we are abbreviating $x_j^i$ by $e$ and $v_j^i$ by $\dot{e}$.

This gives:

$$dx = dg\, x_0 + g\, dx_0, \tag{5.37}$$
$$dv = d\dot{g}\, x_0 + \dot{g}\, dx_0 + dg\, v_0 + g\, dv_0, \tag{5.38}$$
$$de = dg\, e_0 + g\, de_0, \tag{5.39}$$
$$d\dot{e} = d\dot{g}\, e_0 + \dot{g}\, de_0 + dg\, \dot{e}_0 + g\, d\dot{e}_0. \tag{5.40}$$

Substitution into (5.33) gives:

$$M_i^j = F_i x_0^j + p_i v_0^j + \tau_i^k [x_k^j]_0 + S_i^k [v_k^j]_0, \tag{5.41}$$
$$L_i^j = p_i x_0^j + S_i^k [x_k^j]_0, \tag{5.42}$$
$$F_{0i} = F_k g_i^k + p_k \dot{g}_i^k, \tag{5.43}$$
$$p_{0i} = p_k g_i^k, \tag{5.44}$$
$$[\tau_i^j]_0 = \tau_k^j g_i^k + S_k^j \dot{g}_i^k, \tag{5.45}$$
$$[S_i^j]_0 = [S_k^j g_i^k]_0. \tag{5.46}$$

The equations of motion are obtained from the vanishing of $D^*\phi$, which initially gives the system of equations:

$$M_i^j = \frac{dL_i^j}{dt}, \qquad F_{0i} = \frac{dp_{0i}}{dt}, \qquad [\tau_i^j]_0 = \frac{d[S_i^j]_0}{dt}, \tag{5.47}$$

but when one substitutes the previous set of equations for all of the quantities, one finds that the last two sets of equations in (5.47) become:

$$F_i = \frac{dp_i}{dt}, \qquad \tau_i^j = \frac{dS_i^j}{dt}, \tag{5.48}$$

and the first set reduces to an identity.

If one now wishes to define the dynamical state relative to the non-inertial frame then one must make the replacements:

$$dg = g\nabla g, \qquad d\dot{g} = g\nabla \omega, \tag{5.49}$$

in (5.34), along with:

$$dx = g(\nabla g\, x_0 + \nabla x_0), \tag{5.50}$$
$$dv = g(\nabla \omega\, x_0 + \omega \nabla x_0 + \nabla g\, v_0 + \nabla v_0), \tag{5.51}$$
$$de = g(\nabla g\, e_0 + \nabla e_0), \tag{5.52}$$
$$d\dot{e} = g(\nabla \omega\, e_0 + \omega \nabla e_0 + \nabla g\, \dot{e}_0 + \nabla \dot{e}_0), \tag{5.53}$$

in which we have defined:

$$\nabla x_0 = g^{-1}dx, \quad \nabla v_0 = g^{-1}dv, \quad \nabla e_0 = g^{-1}de, \quad \nabla \dot{e}_0 = g^{-1}d\dot{e}. \tag{5.54}$$

We then obtain:

$$\phi_g = (M_i^j + [L_i^j]_0 \omega_j^k)\nabla g_j^i + L_i^j \nabla \omega_j^i \equiv [M_i^k]_0 \nabla g_j^i + [L_i^j]_0 \nabla \omega_j^i, \tag{5.55}$$
$$\phi_0 = \bar{F}_{0i}\nabla x_0^i + [\bar{\tau}_j^i]_0 \nabla[e_j^i]_0 + \bar{p}_{0i}\nabla v_0^i + [\bar{S}_i^j]_0 \nabla[\dot{e}_j^i]_0, \tag{5.56}$$

which makes:

$$[\bar{M}_j^i]_0 = \{ [M_j^i]_0 + F_{0j}x_0^i + p_{0j}v_0^i + [\tau_j^i]_0 + [S_j^k \dot{e}_k^i]_0 \}_{[ij]}, \tag{5.57}$$
$$[\bar{L}_j^i]_0 = \{ [L_j^i]_0 + p_{0j}x_0^i + [S_j^i]_0 \}_{[ij]}, \tag{5.58}$$
$$\bar{F}_{0i} = F_{0i} + p_{0j}\omega_i^j, \tag{5.59}$$
$$\bar{p}_{0i} = p_{0i}, \tag{5.60}$$
$$[\bar{\tau}_j^i]_0 = [\tau_j^i]_0 + [S_k^i]_0 \omega_j^k, \tag{5.61}$$
$$[\bar{S}_k^i]_0 = [S_k^i]_0. \tag{5.62}$$

In these equations, we have set the component matrix of the initial frame $e_0$ equal to $\delta_j^i$, which is no loss of generality. The notation [ij] in the first two expression means that since the matrices $\nabla g_j^i$ and $\nabla \omega_j^i$ are anti-symmetric, as well as $\nabla e_0$ and $\nabla \dot{e}_0$, one must use only the anti-symmetric part of the matrices and tensor products involved:

$$\{F_{0j}x_0^i\}_{[ij]} = \tfrac{1}{2}(F_{0j}x_0^i - F_{0i}x_0^j), \tag{5.63}$$
$$\{p_{0j}x_0^i\}_{[ij]} = \tfrac{1}{2}(p_{0j}x_0^i - p_{0i}x_0^j), \tag{5.64}$$
$$\{p_{0j}v_0^i\}_{[ij]} = \tfrac{1}{2}(p_{0j}v_0^i - p_{0i}v_0^j). \tag{5.65}$$

Of course, when one uses the most common constitutive law for the linear momentum of a point mass – viz., $p_i = m\delta_{ij}v^j$ – the last expression vanishes, but the work of Weyssenhoff [8] on relativistic spinning fluids suggests that in some cases, the presence of a "transverse momentum" contribution might play an important role.

The equations of motion initially take the form:

$$[\bar{M}_j^i]_0 = \nabla_t[\bar{L}_j^i]_0, \qquad \bar{F}_{0i} = \frac{d\bar{p}_{0i}}{dt}, \qquad [\bar{\tau}_j^i]_0 = \frac{d[\bar{S}_j^i]_0}{dt}, \tag{5.66}$$

Using equations (5.59)-(5.62), we first find that the second two take the form:

$$F_{0i} = \nabla_t p_{0i}, \qquad [\tau_j^i]_0 = \nabla_t[S_j^i]_0, \tag{5.67}$$

and substitution of these into the first set of (5.66) leaves only:

$$[M^i_j]_0 = \nabla_t [L^i_j]_0. \tag{5.68}$$

Since this contribution vanished in the inertial frame, one sees that it is essentially a "fictitious" contribution that comes about as a result of the transformation to a non-inertial one.

### 5.2.5 Generalized Noether currents

Although the forces and torques that act on a rigid body can very well be non-conservative, such as viscous drag forces and moments for a rigid body moving in a viscous fluid, nonetheless, the linear and angular momenta are generally assumed to be based in the exterior derivative of a total kinetic energy function:

$$T(v^i, \omega^i_j) = \tfrac{1}{2} p(v) + \tfrac{1}{2} L(\omega), \tag{5.69}$$

with linear and angular momenta:

$$p_i = m \delta_{ij} v^j, \qquad L^j_i = I^{jl}_{ik} \omega^k_l, \tag{5.70}$$

in which $m$ is the mass and $I^{jl}_{ik}$ is the moment of inertia, both of which are assumed constant. Of course, it is more traditional to regard the moment of inertia tensor as having two indices, not four, but that is because one can just as easily represent the elements of the three-dimensional Lie algebra $\mathfrak{so}(3)$ by singly indexed components, such as $\dot\omega^i$, as by doubly indexed matrices; as we saw, the matrix $\omega^i_j$ is the matrix of the adjoint map for the vector $\dot\omega^i$. However, in the present form it becomes more obvious that one is not dealing with the vector space $\mathbf{R}^3$ in both cases, but the Lie algebras of $\mathbf{R}^3$ and $\mathfrak{so}(3)$, which only incidentally have the same dimensions as vector spaces.

Since $\mathcal{O} = [t_0, t_1]$ is one-dimensional again, the stress-energy-momentum tensor reduces to the scalar $T$ of total kinetic energy.

Under a variation $\delta s \in \mathfrak{X}([t_0, t_1]; SO(M))$, whose local form is:

$$\delta s = \delta t \frac{\partial}{\partial t} + \delta x^i \frac{\partial}{\partial x^i} + \delta e^i_j \frac{\partial}{\partial e^i_j}, \tag{5.71}$$

with:
$$\delta x^i = v^i \, \delta t, \qquad \delta e^i_j = \dot e^i_j \delta t, \tag{5.72}$$

then goes to the vector field on $[t_0, t_1]$:

$$J(\delta s) = T \, \delta t. \tag{5.73}$$

The balance principle (4.16) then again takes the form:

$$\frac{dT}{dt} = P, \tag{5.74}$$

in which the total power $P$ delivered to or dissipated by the rigid body by the forces and torques that act on it now takes the form:

$$P = F_i v^i + \tau^i_j \dot{e}^j_i = F_{0i} v^i_0 + [\tau^j_i]_0 \omega^i_j. \tag{5.75}$$

### 5.3    Moving deformable body

Although it would rapidly expand the size of this article to go into all of the details about how the aforementioned generalization of Noether's theorem applies to moving deformable bodies, nevertheless, since we already set up much of the basic machinery in the context of point mechanics, we shall at least comment on what changes when one expands from a spatial body $B$ that is zero-dimensional to one of dimension greater than zero. For instance, one can consider bounded strings (i.e., filaments), compact surfaces, such as membranes, and solid compact objects, which correspond to $B$ having dimensions one, two, and three, respectively.

The next thing that changes, beyond the dimension $\mathcal{O}$, is the number of parameter derivatives, which will then be partial derivatives, instead of a total time derivative. However, partial derivatives with respect to the spatial parameters $a^i$, $i = 1, 2, \ldots, p$ will have the same mathematical status as the partial derivative with respect to the time parameter $a^0 = \tau$, which can lead to a possible confusion of units, unless the time parameter has the same unit as the spatial parameters.

We point out that although the matrix $x^\mu_{,i} = \partial x^\mu / \partial a^i$ does not directly represent the strain state of the body $x: \mathcal{O} \to M$, where $\mathcal{O} = [\tau_0, \tau_1] \times B$, in the usual Cauchy-Green sense of strain, it does, however, allow one to derive that tensor field when one is given two such embeddings $x$, $\bar{x}$ of $\mathcal{O}$. Since they are both embeddings, the composed map $y = \bar{x} \cdot x^{-1}: x(\mathcal{O}) \to M$, which is defined only on the image $x(\mathcal{O})$, is a diffeomorphism onto its own image in $M$. If we assume that $M$ has a metric $g$ defined on it then the Cauchy-Green way of characterizing finite strain (in the *Lagrangian* picture – i.e., relative to the initial state $x(\mathcal{O})$) is to pull $g$ back to each $x \in x(\mathcal{O})$ by way of $y$ and then subtract the value of $g$ at $x$:

$$E = y^* g - g. \tag{5.76}$$

The reader that is comfortable with the basic concerns of differential geometry will already notice that, in effect, one is really comparing the deformed metric at one point to the undeformed metric at a distinct point, which sounds geometrically suspicious in the absence of a connection that would allow one to identify the tangent spaces at distinct points. Furthermore, that identification can become path-dependent when there is more than one geodesic between the points, such as conjugate points. The usual way that one resolves this situation in non-relativistic continuum mechanics is to deal with only *M* in the form of a Euclidian space, so it is reasonable to say that *g* is "the same" everywhere. Similarly, one mostly deals with infinitesimal strain in practice, so one sees that classical, non-relativistic continuum mechanics breaks down for manifolds in which the curvature of the metric becomes appreciable – perhaps, in the vicinity of dense astronomical objects, such as neutron stars and black holes – and the extension from infinitesimal strains to finite ones.

The local expression for the components of *E* takes the form:

$$E_{\mu\nu}(x) = y^{\kappa}_{,\mu}(x) y^{\lambda}_{,\nu}(x) g_{\kappa\lambda}(y(x)) - g_{\mu\nu}(x). \tag{5.77}$$

When $M = \mathbf{R}^m$, one can associate each diffeomorphism *y* with a unique *displacement vector field*:

$$\mathbf{u}(x) = y(x) - x = [y^{\mu}(x) - x^{\mu}]\partial_{\mu}. \tag{5.78}$$

The converse statement is not true, though; the displacement vector field $u^{\mu}(x) = -x^{\mu}$ will take every $x^{\mu}$ to the same point, namely, the origin.

One can then see that $y^{\mu}_{,\nu} = \delta^{\mu}_{\nu} + u^{\mu}_{,\nu}$, which puts $E_{\mu\nu}$ into the form:

$$E_{\mu\nu} = g_{\kappa\mu} u^{\kappa}_{,\nu} + g_{\kappa\nu} u^{\kappa}_{,\mu} + g_{\kappa\lambda} u^{\kappa}_{,\mu} u^{\lambda}_{,\nu}, \tag{5.79}$$

as long as the components of *g* are constant.

It is also possible to pull the initial metric *g* to the deformed state *y* by way of *y*-1, which then corresponds to the Eulerian picture of deformation and strain. Another way of characterizing the strain associated with the deformation of $\mathcal{O}$ from *x* to $\bar{x}$ is to pull *g* back to $\mathcal{O}$ using each embedding and then subtract them to define a strain tensor on $\mathcal{O}$ itself:

$$E_{ij}(\tau, a) = \bar{x}^{\mu}_i \bar{x}^{\nu}_j g_{\mu\nu}(\bar{x}) - x^{\mu}_i x^{\nu}_j g_{\mu\nu}(x). \tag{5.80}$$

(See Raymer [9] and Schöpf [**10**] for a discussion of general relativistic elasticity.)

If our kinematical state is of the form $s(a) = (a^i, x^{\mu}(a), x^{\mu}_i(a))$ then we should expect our dynamical state to be the pull-down to $\mathcal{O}$ by s of a vertical 1-form on $J^1(\mathcal{O}; M)$ that locally looks like:

$$\phi = F_{\mu} dx^{\mu} + \Pi^i_{\mu} dx^{\mu}_i, \tag{5.81}$$

in which the functional dependency of the components on the coordinates of $J^1(\mathcal{O}; M)$ embodies the constitutive laws for *B*.

Of course, just as the generalized velocity matrix $x^{\mu}_i$ matrix is only indirectly related to the strain tensor, similarly, the generalized momentum matrix $\Pi^i_{\mu}$ is distantly related to the energy-momentum-stress tensor that one usually deals with. We shall not go into further details, except to say that the equations of motion for a moving deformable body that result from the zeroes of the first-variation functional:

$$F_\mu = \frac{\partial \Pi^i_\mu}{\partial a^i} \tag{5.82}$$

are also related, but not identical, to the usual equations of elasticity or fluid motion in the same way.

When one expands the dimension of $B$ beyond zero, one opens up the possibility that $\mathcal{O}$ will have more transformations acting on it than just the uniform time translations that act on $[\tau_0, \tau_1]$. Under the Noether map, the stress-energy momentum tensor $T^i_j$ will then no longer be simply the 1×1 matrix of kinetic energy, but will be a $p \times p$ matrix. Hence, the non-conservation of the resulting current $J^i = T^i_j \delta x^j$ for a divergenceless variation $\delta x^j$ will have contributions from not only the power delivered/dissipated by the motion of $B$, but also the spatial gradients of its momentum and stress.

It is even possible to extend the previous introduction of moving frames to the case of extended bodies and still be dealing with physically meaningful generalities. In fact, this is the essence of the Cosserat [**11**] approach to the mechanics of deformable bodies, in which one considers not only internal stresses acting upon the body, but also internal stress moments or force-couples. (More recent discussions of the mechanics of Cosserat media can be found in Teodorescu [**12**] and the IUTAM Conference Proceedings [**13**].)

## 6     Summary

Despite the length of this paper, the basic conclusion is simple enough to state: When one bases the definition of an extremal on the zeroes of the first-variation functional, instead of the critical points of an action functional, the enlargement of scope to include non-conservative mechanical systems also results in an enlargement of the scope of Noether's theorem to associate balance principles for non-conserved currents with symmetries of the first-variation functional.

In the simplest case of a pointlike mass moving under the influence of a non-conservative external force, the only symmetry that one can consider is time-translation, its associated non-conserved current is kinetic energy, and the balance principle couples the time derivative of the kinetic energy with the power delivered to or dissipated by the external force. Similarly, when one considers the rigid body, which is essentially a moving orthonormal frame, instead of a moving point, the only thing that changes is that the power delivered/dissipated comes from two contributions that are due to non-conserved external forces and torques.

When one goes to extended matter distributions, one finds that even the (energy-)momentum density 1-form might possibly be inexact, along with the force 1-form, since even for an irrotational covelocity 1-form $u$ can produce an inexact momentum density 1-form $\rho u$ when the gradient of the mass density $\rho$ is not collinear with the velocity. Similarly, one might even consider a Cosserat approach to the description of the kinematical state of the extended body that would include the possibility of internal stress moments that act on the orthonormal frames at each point of it, in addition to the internal stresses, external forces, and moments.

**References**

1. D. H. Delphenich, Ann. Phys. (Berlin) **18** (2009), 649-670.
2. D. H. Delphenich, Ann. Phys. (Berlin) **18** (2009), 45-56.
3. C. Lanczos, *The Variational Principles of Mechanics, 4$^{th}$ ed.* (Dover, New York, 1986); republication of a 1970 edition that was published by the University of Toronto Press as volume 4 of their *Mathematical Expositions;* the first edition was published in 1949.
4. H. Weyl, Ann. Math. (N.Y.) **36** (1935), pp. 607-629.
5. H. Rund, *The Differential Geometry of Finsler Spaces* (Springer, Berlin, 1959).
6. D. Bao, S.-S. Chern, Z. Shen, *An Introduction to Riemann-Finsler Geometry* (Springer, New York, 2000).
7. V. I. Arnol'd, *Mathematical Methods in Classical Mechanics* (Springer, Berlin, 1978).
8. J. Weyssenhoff and A. Raabe, Acta Phys. Pol. **9** (1947), pp. 7-18, 19-25, 26-33, 34-45, 46-53.
9. C. B. Raymer, Proc. Roy. Soc. London **2724** (1963), 44-53.
10. H.-G. Schöpf, Ann. Phys. (Leipzig) **12** (1964), 377-395.
11. E. Cosserat and F. Cosserat, *Théorie des corps déformables* (Hermann, Paris, 1909).
12. C. Teodorescu, *Dynamics of Linear Elastic Bodies* (Abacus Press, Tunbridge Wells, 1972).
13. E. Kröner, ed., *Mechanics of Generalized Continua,* Proc. of 1967 IUTAM Symposium in Freudenstadt and Stuttgart (Springer, Berlin, 1968).